\documentclass[11pt]{article}
\usepackage[utf8]{inputenc}
\usepackage[T1]{fontenc}
\usepackage{authblk}     
\usepackage{hyperref}    
\usepackage{graphicx}    
\usepackage{enumitem}    
\usepackage{amsmath}     
\usepackage{cite}        
\usepackage{setspace}
\usepackage[labelfont=bf, labelsep=period]{caption} 

\newcommand{\figcap}[2]{%
    \caption[#1]{\textbf{#1.} #2}%
}

\title{\textbf{Development of an Agentic AI Model for NGS Downstream Analysis Targeting Researchers with Limited Biological Background}}

\author[1,2]{Donghyeon Lee}
\author[1,2]{Dongseok Kim}
\author[3]{Seokhwan Ko}
\author[1,2, *]{Seo-Young Park}
\author[3, *]{Junghwan Cho}

\affil[1]{Department of Biomedical Science, Kyungpook National University, Daegu, Republic of Korea}
\affil[2]{BK21 Four Program, School of Medicine, Kyungpook National University, Daegu, Republic of Korea}
\affil[3]{Clinical Omics Institute, Kyungpook National University, Daegu, Republic of Korea}

\date{} 

\begin{document}

\maketitle

\renewcommand{\thefootnote}{\fnsymbol{footnote}}
\footnotetext[1]{Corresponding authors: seoyoungyuri.park@gmail.com (S.-Y. Park), cho.junghwan@gmail.com (J. Cho)}
\renewcommand{\thefootnote}{\arabic{footnote}}

\begin{figure}[htbp]
    \centering
    \includegraphics[width=1\textwidth]{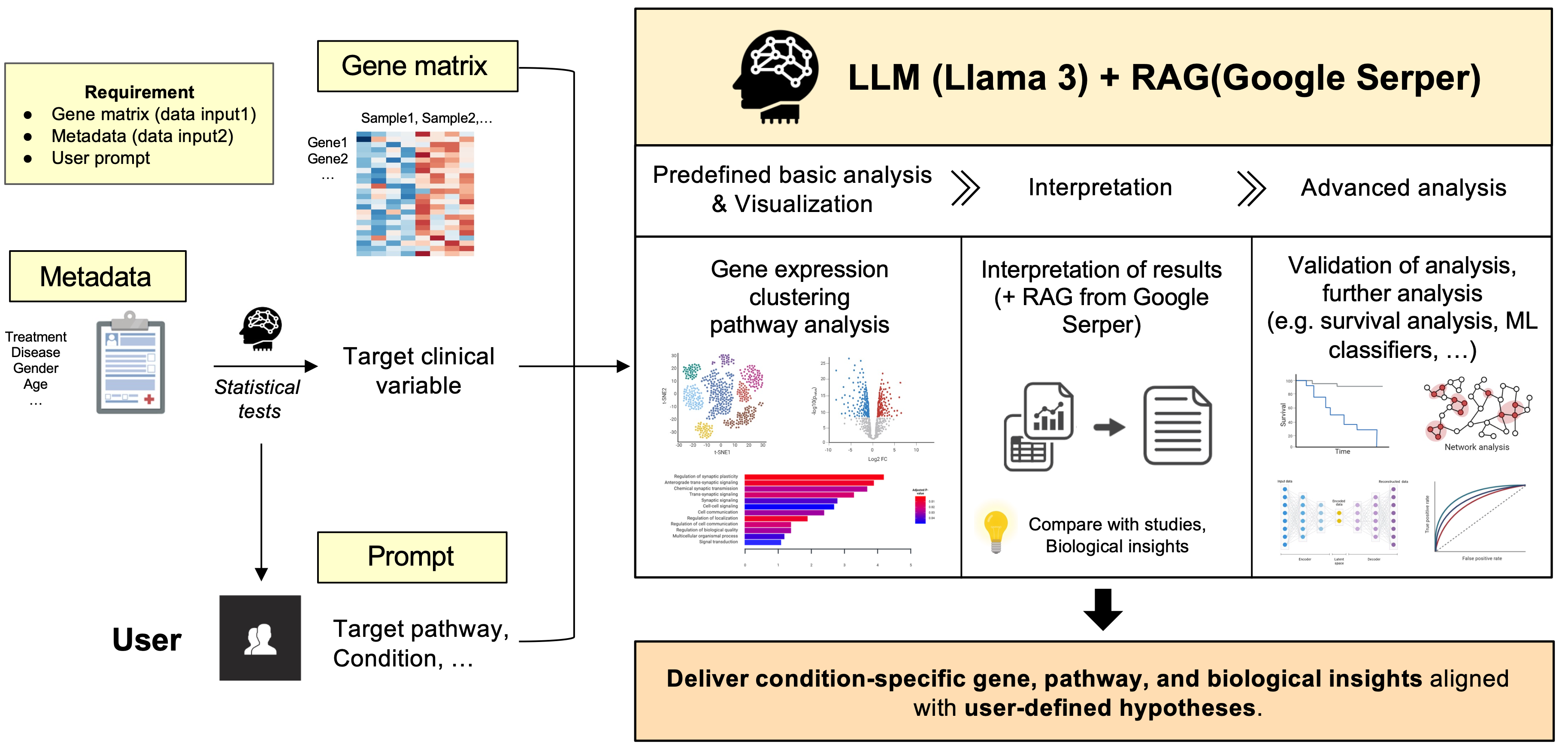}
    
    \figcap{Overview of the Proposed Agentic AI Framework}
    {The pipeline integrates a gene expression matrix, clinical metadata, and user-defined prompts to generate condition-specific biological insights. Target clinical variables for comparison are first selected from the metadata via statistical testing and presented to the user. Using these results, the user formulates a prompt that defines the clinical comparison of interest and relevant target pathways or conditions, which then guides predefined analyses such as differential expression, clustering, and pathway enrichment. The resulting outputs are subsequently interpreted by an LLM (Llama 3 70B) with retrieval-augmented generation using Google Serper, enabling contextualization of findings based on existing literature. As advanced analysis, additional procedures including survival analysis and machine-learning classifiers provide further validation and hypothesis testing. The overall framework delivers gene- and pathway-level interpretations and insights explicitly aligned with user-specified hypotheses and clinical conditions.}
    
    \label{fig:overview}
\end{figure}


\section*{1. INTRODUCTION}
Next-Generation Sequencing (NGS) has revolutionized genomic research, enabling high-throughput analysis of gene expression data. However, downstream analysis such as differential expression gene (DEG) identification, clustering, and pathway analysis requires substantial biological background knowledge and computational expertise. Researchers lacking this expertise often face barriers in interpreting results and extending analyses to advanced stages like predictive modeling or hypothesis generation. 

Currently, numerous studies leverage agentic AI to process RNA-seq data; for example, a variety of agents such as CellAgent, BioAgents, and CompBioAgent have been proposed \cite{xiao2024, mehandru2025, zhang2025}. These AI agents assist with preprocessing and analysis for either scRNA-seq and bulk RNA-seq data, and they perform standard downstream analyses, such as trajectory analysis and clustering. However, existing studies using Agentic AI did not provide specialized downstream analysis, focusing instead on general workflows without tailored interpretations for novices. 

To address this, we developed an Agentic AI model that automates NGS downstream analysis, provides biological interpretations, and recommends/executes advanced methods. Through this process, users can compare their results with findings from previous studies, obtain meaningful insights, generate and test hypotheses, and receive guidance on which further analyses to perform for validation and additional exploration, thereby specifically supporting downstream analysis. Built on Llama 3 70B for efficient reasoning and RAG for literature-backed insights, the model is deployed as an interactive Streamlit web app, making it accessible to newcomers in bioinformatics.

The model's novelty lies in its agentic nature: it autonomously plans, executes, and reflects on analyses, using user-provided gene expression matrices and clinical data. The model is designed to execute and present prior statistical tests for each clinical variable and accept user prompts specifying the perspective from which the data should be analyzed. The model integrates bioinformatics tools (Biopython \cite{cock2009}, GSEApy \cite{fang2023}, gProfiler \cite{kolberg2023}) with LLM-driven code generation, allowing seamless transition from basic to advanced analyses. This extends prior works like GenoMAS \cite{liu2025} for transcriptomic analysis and AutoBA \cite{zhou2024} for multi-agent omics, by incorporating adaptive RAG-based literature search and user-selected advanced workflows.

\section*{2. MATERIALS AND METHODS}

\subsection*{System Overview}
The model follows a modular, agentic workflow implemented in Python, with Streamlit providing an interactive user interface. Users can upload their own gene expression matrices (RNA-seq data) and clinical information in CSV format or alternatively use example data consisting of $200 \text{ genes} \times 100 \text{ samples}$ enriched for cancer-related genes. Genes are extracted to ensure consistent handling of gene identifiers across all components of the pipeline, and the expression matrix is normalized using a $\log_2$ transformation to stabilize variance and reduce the impact of extreme values. To relate clinical features to the user-defined hypothesis, the system computes a statistical significance table for the clinical data using Pearson correlation or t-tests between hypothesis-related variables and other clinical columns, summarizing the data type, test used, test statistic, and $p$-value in a single table.

\subsection*{Core Downstream Analysis and Visualization}
Based on the processed inputs and clinical variables, the model performs core downstream analyses commonly used in RNA-seq workflows. For DEG analysis, a t-test is applied to compare groups defined by a user-selected clinical variable, and log fold changes (logFC) are calculated to quantify expression differences between groups. Unsupervised structure in the data is explored via k-means clustering on the transposed expression matrix, with clusters visualized using principal component analysis (PCA). To connect DEGs to biological processes, pathway analysis is conducted using gProfiler (0.3.5) and gene set enrichment analysis (GSEA) through GSEA.py (1.1.11) on significantly altered genes, and the top enriched pathways are summarized in bar plots. The model generates multiple visual outputs, including a volcano plot annotated with the top 10 genes, a PCA plot colored by the selected clinical variable and accompanied by a legend, and functional bar plots highlighting key pathways, thereby offering both statistical and biological views of the data.

\subsection*{Interpretation, Literature Integration, and Advanced Agentic Actions}
To support users with limited bioinformatics or biology background, the model employs an LLM (Llama 3.3-70b) to interpret DEG, clustering, and pathway results. The model further integrates a literature-aware RAG pipeline: PubMed is queried via Entrez using terms derived from dysregulated genes and enriched pathways, abstracts are embedded with sentence-transformer models and stored in a FAISS vector database for retrieval and summarization. Building on these insights, the LLM recommends three advanced analysis methods tailored to the specific dataset and findings, each accompanied by a description, rationale, and executable Python code. Users can select one of these methods via a Streamlit selectbox and trigger execution with a ``run'' button directly in the web interface. The LLM then summarizes the advanced analysis outputs for beginners and generates downloadable reports. Throughout, the workflow is organized into interactive stages with dedicated buttons, preserving user control while maintaining an agentic, autonomous orchestration of data processing, analysis, interpretation, and method recommendation.

\section*{3. RESULTS}
Using example data, the model identifies $\sim 40$ significant DEGs ($p < 0.05$), clusters samples into two groups correlated with \texttt{Disease\_Status}, and simulates pathway IDs. Volcano plot highlights top DEGs (e.g., \textit{Gene1} with high $-\log_{10}(p\text{-value})$). PCA plot color points by selected variables (e.g., \texttt{Disease\_Status}), revealing group separation. Clinical significance table shows $p$-values (e.g., \texttt{Disease\_Status} vs. Age: $p = 0.3$, not significant).

Based on genes obtained from basic analysis (e.g., upregulated oncogenes like \textit{BRAF} in diseased samples), the LLM derives insights such as potential immune pathway dysregulation, reasoned by $\text{logFC} > 1$ in DEG results indicating tumor progression. RAG summaries from PubMed abstracts provide literature-backed evidence, e.g., \textit{BRAF} mutations linked to poor prognosis in lung cancer (logical ground: correlation with clinical survival data, $p < 0.05$). Advanced recommendations include survival modeling (code: \texttt{statsmodels} CoxPHFitter), executed on user selection, yielding hazard ratios grounded in DEG-driven hypothesis testing.

\section*{4. CONCLUSION}
This Agentic AI model democratizes NGS downstream analysis by automating complex tasks and providing interpretable insights for non-experts. By integrating LLM for code generation and RAG for evidence-based recommendations, it reduces barriers in bioinformatics. Future extensions could include gene network-based drug response prediction for target discovery and multi-omics support. The Streamlit app ensures accessibility, potentially accelerating discovery in precision medicine.

\section*{ACKNOWLEDGEMENT}
This work was supported by the Brain Pool Program through the National Research Foundation of Korea (NRF), funded by the Ministry of Science and ICT (RS-2023-00283791), the Ministry of Education, Korea (2021R1I1A3056903), by Basic Science Research Program through the NRF funded by the Ministry of Education (RS-2024-00459836), and by a grant of the Korea Health Technology R\&D Project through the Korea Health Industry Development Institute (KHIDI), funded by the Ministry of Health \& Welfare, Republic of Korea (RS-2025-02263414). The corresponding authors (S.-Y. Park and J. Cho) declare conflicts of interest related to ongoing patent applications associated with this research.


\end{document}